\documentclass[a4paper]{report}
\usepackage[utf8]{inputenc}
\usepackage[T1]{fontenc}
\usepackage{RJournal}
\usepackage{amsmath,amssymb,array}
\usepackage{booktabs}

\providecommand{\tightlist}{%
  \setlength{\itemsep}{0pt}\setlength{\parskip}{0pt}}

\usepackage{longtable}

\newlength{\cslhangindent}
\newlength{\csllabelwidth}
\newlength{\cslentryspacingunit} 
  {}%
  {\par}
 {
  \setlength{\parindent}{0pt}
  \ifodd #1
  \let\oldpar\par
  \def\par{\hangindent=\cslhangindent\oldpar}
  \fi
  \setlength{\parskip}{#2\cslentryspacingunit}
 }%
 {}
\usepackage{calc}

\usepackage{amsmath} \usepackage{amssymb} \usepackage[ruled,noline]{algorithm2e} \usepackage{amsfonts} \usepackage{xcolor} \usepackage{tikz}

\begin{document}

\sectionhead{Contributed research article}
\volume{XX}
\volnumber{YY}
\year{2025}
\month{May}

\begin{article}

\title{panelPomp: Analysis of Panel Data via Partially Observed Markov Processes in R}

\author{by Carles Bretó, Jesse Wheeler, Aaron A. King, and Edward L. Ionides}

\maketitle

\abstract{%
Panel data arise when time series measurements are collected from multiple, dynamically independent but structurally related systems. Each system's time series can be modeled as a partially observed Markov process (POMP), and the ensemble of these models is called a PanelPOMP. If the time series are relatively short, statistical inference for each time series must draw information from across the entire panel. The component systems in the panel are called units; model parameters may be shared between units or may be unit-specific. Differences between units may be of direct inferential interest or may be a nuisance for studying the commonalities. The R package panelPomp supports analysis of panel data via a general class of PanelPOMP models. This includes a suite of tools for manipulation of models and data that take advantage of the panel structure. The panelPomp package currently highlights recent advances enabling likelihood based inference via simulation based algorithms. However, the general framework provided by panelPomp supports development of additional, new inference methodology for panel data.
}

\hypertarget{introduction}{%
\section{Introduction}\label{introduction}}

Collections of time series, known as panel data or longitudinal data, are commonplace across the sciences, medicine, engineering and business.
Examples include biomarkers measured over time across a panel of patients in a clinical trial \citep{ranjeva17, ranjeva19},
ecological predator-prey dynamics for lake plankton measured within a season across a panel of years or lakes \citep{marino19},
and interactions between self driving cars and pedestrians \citep{domeyer22}.
Generically, we say each time series is associated with measurements on a unit.
Not infrequently, the time series data available on a single unit are too short or too noisy to adequately identify parameters of interest, and yet large collections of such time series analyzed jointly may suffice.
This paper presents software enabling the fitting of general nonlinear nonstationary partially observed stochastic dynamic models to panel data.
The methodology provides flexibility in model specification, giving the user considerable freedom to develop models appropriate for the system under study.

Scientific motivations to fit a panel of partially observed Markov processes (PanelPOMP) model to panel data \citep{breto20} are similar to the motivations to fit a partially observed Markov process (POMP) model to time series data from a single unit.
POMP models provide a general framework for mechanistic modeling of nonlinear dynamic systems \citep{breto09}.
However, relatively few of the many methodologies developed for time series analysis via POMPs have been extended to panel analysis.
The larger datasets and higher-dimensional parameter spaces arising in PanelPOMPs have challenged the Monte Carlo methods that have proved successful for nonlinear time series.
Our \CRANpkg{panelPomp} package takes advantage of newly developed methodology for PanelPOMP models, as well as building a foundation on which additional methodology for this class of models can be built and tested.

Existing widely-used panel methodology has built on the linear Gaussian model \citep{croissant08}.
Panel analysis of generalized linear models with dependence can be carried out using generalized estimating equations \citep{halekoh06}.
By contrast, \CRANpkg{panelPomp} is designed to facilitate analysis using arbitrary PanelPOMP models, with dynamic relationships and dependence on covariate processes specified according to scientific considerations rather than the constraints of statistical software.
Beyond supplying implementations of inference algorithms for PanelPOMP models, the \CRANpkg{panelPomp} package provides a framework for developing and sharing new models and methods.
This is done by providing a way of writing basic mathematical functions that comprise a general PanelPOMP model as easy to write C-code snippets \citep{king16}, which makes the \CRANpkg{panelPomp} package both computationally efficient and readily adaptable to new developments in methodology for PanelPOMP models.

Analysis of experimental or observational studies often assumes independent outcomes for units given their treatment, and here we take the same approach for panel models.
That is, the underlying dynamic process that is used to define a PanelPOMP model is assumed be independent across units.
A collection of POMP models with dynamic dependence between units is called a SpatPOMP, a name motivated by the dependence between nearby locations in spatiotemporal models.
Software for general SpatPOMP models is available in the \CRANpkg{spatPomp} package \citep{asfaw24}.
Addressing SpatPOMP models introduces complexities in both software and methodologies that can be avoided when the dynamic processes are independent.
The goal of the \CRANpkg{panelPomp} package is to take full advantage of the independence inherent in the PanelPOMP model structure.

\hypertarget{statistical-models}{%
\section{Statistical models}\label{statistical-models}}

The general scope of the \CRANpkg{panelPomp} package requires notation concerning random variables and their densities in arbitrary spaces.
The notation below allows us to talk about these things using the language of mathematics, enabling precise description of models and algorithms.

Units of the panel can be identified with numeric labels \(\{1,2,\dots,U\}\), which we also write as \(1\,{:}\,U\).
Let \(N_u\) be the number of measurements collected on unit \(u\), and write the data as \(y^*_{u,1:N_{u}} = \{y^*_{u,1},\dots,y^*_{u,N_u}\}\) where \(y^*_{u,n}\) is collected at time \(t_{u,n}\) with \(t_{u,1}<t_{u,2}<\dots<t_{u,N_u}\).
The data are modeled as a realization of an observable stochastic process \(Y_{u,1:N_u}\) which is dependent on a latent Markov process \(\{X_u(t),t_{u,0}\le t\le t_{u,N_u}\}\) defined subsequent to an initial time \(t_{u,0}\le t_{u,1}\).
Requiring that \(\{X_u(t)\}\) and \(\{Y_{u,i},i\neq n\}\) are independent of \(Y_{u,n}\) given \(X_u(t_{u,n})\), for each \(n\in 1\,{:}\, N_{u}\), completes the partially observed Markov process (POMP) model structure for unit \(u\).
For a PanelPOMP we require additionally that all units are modeled as independent.

The latent process at the observation times is written \(X_{u,n}=X_u(t_{u,n})\).
We suppose that \(X_{u,n}\) and \(Y_{u,n}\) take values in arbitrary spaces \(\mathbb{X}_{u}\) and \(\mathbb{Y}_{u}\) respectively.
Using the independence of units, conditional independence of the observable random variables, and the Markov property of the latent states, the joint distribution of the entire collection of latent variables \(\mathbf{X} = \{X_{u,0:N_u}\}_{u = 1}^U\) and observable variables \(\mathbf{Y} = \{Y_{u,1:N_u}\}_{u = 1}^U\) can be written as
\[
f_{\mathbf{X}\mathbf{Y}}(\mathbf{x}, \mathbf{y}) = \prod_{u = 1}^U f_{X_{u, 0}}(x_{u,0}; \theta)\prod_{n = 1}^{N_u} f_{Y_{u, n}|X_{u, n}}(y_{u, n}|x_{u, n}; \theta)f_{X_{u, n}|X_{u, n-1}}(x_{u, n}|x_{u, n-1}; \theta),
\]
where \(\theta\in\mathbb{R}^{D}\) is a parameter vector.
This representation is useful as it demonstrates how any PanelPOMP model can be fully described using three primary components: the unit transition densities
\(f_{X_{u,n}|X_{u,n-1}}(x_{u,n}\,|\, x_{u,n-1}\,;\,\theta)\),
measurement densities
\(f_{Y_{u,n}|X_{u,n}}(y_{u,n}\,|\, x_{u,n}\,;\,\theta)\),
and initialization densities
\(f_{X_{u, 0}}(x_{u,0}; \theta)\).
Each class of densities are permitted to depend arbitrarily on \(u\) and \(n\), allowing non-stationary models and the inclusion of covariate time series.
In addition to continuous-time dynamics, the framework includes discrete-time dynamic models by specifying \(X_{u,0:N_u}\) directly without ever defining \(\{X_u(t),t_{u,0}\le t\le t_{u,N_u}\}\).
We also permit the possibility that some parameters may affect only a subset of units, so that the parameter vector can be written as
\(\theta=(\phi,\psi_1,\dots,\psi_U)\),
where the densities described above can be written as
\begin{eqnarray}
f_{X_{u,n}\vert X_{u,n-1}}(x_{u,n}\,|\, x_{u,n-1} \,;\, \theta)
&=&
f_{X_{u,n}|X_{u,n-1}}(x_{u,n}\,|\, x_{u,n-1} \,;\, \phi,\psi_u) \label{eq:proc} 
\\
f_{Y_{u,n}|X_{u,n}}(y_{u,n}\,|\, x_{u,n} \,;\, \theta) &=& f_{Y_{u,n}|X_{u,n}}(y_{u,n}\,|\, x_{u,n} \,;\, \phi,\psi_u) \label{eq:meas} 
\\
f_{X_{u,0}}(x_{u,0} \,;\, \theta) &=& f_{X_{u,0}}(x_{u,0} \,;\, \phi,\psi_u). \label{eq:init} 
\end{eqnarray}
Then, \(\psi_{u}\) is a vector of \emph{unit-specific} parameters for unit \(u\), and \(\phi\) is a \emph{shared} parameter vector.
We suppose \(\phi\in\mathbb{R}^{A}\) and \(\psi_{u}\in\mathbb{R}^{B}\), so the dimension of the parameter vector \(\theta\) is \(D=A+BU\).
In practice, the densities in Eqs. \eqref{eq:proc}--\eqref{eq:init} serve two primary roles in PanelPOMP models: evaluation and simulation.
Each function can depend on the unit to which it belongs, and therefore are specified in the unit-specific POMP models that together define the PanelPOMP.
This feature is reflected in the software implementation, where the fundamental tasks of simulation and evaluation are defined in the unit \texttt{pomp} objects, as described in in Table \ref{tab:funs}.

\begin{table}[ht]
\begin{center}
\begin{tabular}{llr}\hline
Method            & Operation & Function\\\hline
\texttt{rprocess} & Simulate from Eq.~(\ref{eq:proc}) & $f_{X_{u,n}\vert X_{u,n-1}}(x_{u,n}\,|\, x_{u,n-1} \,;\, \phi,\psi_u)$\\
\texttt{dprocess} & Evaluate Eq.~(\ref{eq:proc}) & $f_{X_{u,n}\vert X_{u,n-1}}(x_{u,n}\,|\, x_{u,n-1} \,;\, \phi,\psi_u)$\\
\texttt{rmeasure} & Simulate from Eq.~(\ref{eq:meas}) & $f_{Y_{u,n}|X_{u,n}}(y_{u,n}\,|\, x_{u,n} \,;\, \phi,\psi_u)$\\
\texttt{dmeasure} & Evaluate Eq.~(\ref{eq:meas}) & $f_{Y_{u,n}|X_{u,n}}(y_{u,n}\,|\, x_{u,n} \,;\, \phi,\psi_u)$\\
\texttt{rinit}   & Simulate from Eq.~(\ref{eq:init}) & $f_{X_{u,0}}(x_{u,0} \,;\, \phi,\psi_u)$\\
\texttt{dinit}   & Evalutate Eq.~(\ref{eq:init}) & $f_{X_{u,0}}(x_{u,0} \,;\, \phi,\psi_u)$\\\hline
\end{tabular}
\end{center}
\caption{\label{tab:funs}Methods of a unit model in a \texttt{panelPomp} object and their mathematical definitions.}
\end{table}

In addition to the functions listed in in Table \ref{tab:funs}, additional basic mathematical functions can be specified for the unit objects of a \texttt{panelPomp} model as needed.
For instance, functions \texttt{rprior} and \texttt{dprior}---which represent the operations of simulating from and evaluating a prior density function, respectively---can be specified as part of the unit objects if desired.
The current version of the package (\texttt{1.7.0.0}), however, currently emphasizes the likelihood-based, plug-and-play methodologies described in Section 3, which do not require the specification of a prior distribution.

\hypertarget{model-representation-in}{%
\subsection{\texorpdfstring{Model representation in \CRANpkg{panelPomp}}{Model representation in }}\label{model-representation-in}}

The package uses the S4 functional object oriented programming approach in R \citep{wickham19} in order to represent PanelPOMP models as \texttt{panelPomp} objects.
Because a PanelPOMP model comprises multiple POMP models, a \texttt{panelPomp} object is essentially a structured collection of \texttt{pomp} objects from the \CRANpkg{pomp} package \citep{king16}.
The \texttt{panelPomp} class contains three attributes: \texttt{unit\_objects}, a list containing the models for each unit; \texttt{shared}, a named numeric vector of parameters that are shared across all units; and \texttt{specific}, a matrix of parameters that are unique to each unit.

Typically, creating \texttt{panelPomp} objects involves supplying a dataset for each unit and explicitly defining the mathematical functions in Table \ref{tab:funs}.
The functions can be defined either using \texttt{R} functions or C-code snippets, the latter offering the advantage of reduced computing times.
The construction of \texttt{pomp} objects using both \texttt{R} functions and C-code snippets is a topic discussed in detail in \citet{king16} and various online tutorials for the \CRANpkg{pomp} package.
To avoid redundancy with these existing material, we focus instead on the novel features of the \CRANpkg{panelPomp} package.
Specifically, we illustrate how to construct a \texttt{panelPomp} object when a collection of \texttt{pomp} objects is already available.

For demonstration purposes, we consider a stochastic version of the discrete-time Gompertz population model \citep{winsor32}.
This model is a popular choice for representing the exponential growth and decay observed in numerous ecological populations \citep{auger21, smith23, linden09}.
It serves as a useful example due to its nonlinear, non-Gaussian nature, which can be transformed into a linear Gaussian model through log transformations.
Consequently, the model is a popular choice for demonstrating the capabilities of plug-and-play algorithms on a nonlinear, non-Gaussian model where the likelihood can be exactly calculated using linear Gaussian techniques \citep{breto20}.
It further serves as an illustrative case study for \CRANpkg{panelPomp}, as researchers may use similar models to describe populations dynamics at distinct observation sites or experimental treatments.
This approach allows for formal statistical testing to determine whether characteristics of the population dynamics are shared across populations or are unique to each unit.

For each unit \(u\), the latent population density \(X_{u,n}\) at time \(n\) is recursively modeled according to Eq. \eqref{eq:gomp}.
\begin{eqnarray}
X_{u, n+1} = \kappa_u^{1-e^{-r_u}}X^{e^{-r_u}}_{u, n}\epsilon_{u, n}, \label{eq:gomp} 
\end{eqnarray}
where \(\epsilon_{n, u}\) are independent and identically-distributed log-normal random variables with \(\log \epsilon_{u, n}\sim \text{N}(0, \sigma^2_{u})\).
The observed population \(Y_{u, n}\) at time \(n\) is assumed to follow a log-normal distribution, independently and identically distributed, conditioned on the value \(X_{u, n}\),
\[
\log Y_{u, n} | X_{u, n} \sim \text{N}\big(\log X_{u, n}, \tau_u^2\big). \label{eq:gompMeas}
\]

Below, we use the \texttt{pomp::gompertz} constructor function to build these unit-specific models.
These constructors simultaneously build the unit models described above, and stores a simulation from the model in the \texttt{data} attribute.
The \texttt{times} argument in the constructor indicates the observation times for each unit;
these times do not need to be the same when constructing \texttt{panelPomp} objects, as highlighted below.

\begin{verbatim}
gomp_u1 <- pomp::gompertz(
  K = 1, r = 0.1, sigma = 0.1, tau = 0.1, X_0 = 1, times = 1:20, seed = 111
)

gomp_u2 <- pomp::gompertz(
  K = 1.5, r = 0.1, sigma = 0.1, tau = 0.07, X_0 = 1, times = 1:21, seed = 222
)

gomp_u3 <- pomp::gompertz(
  K = 1.2, r = 0.1, sigma = 0.1, tau = 0.15, X_0 = 1, times = 3:25, seed = 333
)
\end{verbatim}

In this example, we construct three unit objects, \texttt{gomp\_u1}, \texttt{gomp\_u2}, and \texttt{gomp\_u3}, each of which is a \texttt{pomp} object.
Each object contain a vector of parameters \texttt{K}, \texttt{r}, \texttt{sigma} \texttt{tau}, and \texttt{X\_0} that correspond to the parameters \(\kappa_u, r_u, \sigma_u, \tau_u\) and \(X_{u, 0}\), respectively.
To build a \texttt{panelPomp} object, we pass a list of these models and specify which parameters are \texttt{shared} across units and which are \texttt{specific} to each unit.

\begin{verbatim}
gomp <- panelPomp(
  object = list(gomp_u1, gomp_u2, gomp_u3),
  shared = c("r" = 0.1, "sigma" = 0.1),
  specific = c("K", "tau", "X_0")
)
\end{verbatim}

The parameter values of the \texttt{shared} vector need to be explicitly defined, replacing the separate values in the parameter vectors for each unit object.
The unit-specific parameter are extracted from the constituent \texttt{pomp} objects.
In the above construction, we are creating a \texttt{panelPomp} model that has shared parameters \texttt{r} and \texttt{sigma}.
Mathematically, this is equivalent to assuming that, for all \(u \in 1:U\), \(r_u = r = 0.1\) and \(\sigma_u = \sigma = 0.1\).
The choice of which parameters are shared and which are unit-specific can be modified, as the helper function described in the following subsection enable changing the original parameter specifications.

Several pre-built models are also included in the package via constructor functions, including:

\begin{itemize}
\tightlist
\item
  \texttt{contacts()} creates a dynamic model for the variation in sexual contacts for the data of \citet{vittinghoff99}.
  The model supposes each individual has a latent rate of making sexual contacts that evolves overtime, allowing for heterogeneity between and within individuals, auto-correlation in individual rates over time, and a trend in rates over the time of the study \citep{romero-severson15}.
\item
  \texttt{panelMeasles()} creates a PanelPOMP model for measles incidence data in several UK cities, based on the model of \citet{he10}.
  This model is a stochastic compartmental model that describes Measles incidence data using a susceptible, exposed, infected, recovered (SEIR) model.
  The source code provides a useful demonstration of how users can build compartmental models for panel data from an infectious disease outbreak.
\item
  \texttt{panelRandomWalk()} constructs a collection of independent latent Gaussian random walk models. After building the panel model, the constructor populates the \texttt{data} slots for the created \texttt{unitObjects} by simulating from the created model.
\item
  \texttt{panelGompertz()} creates a collection of the stochastic Gompertz population models described in this section. The \texttt{data} slots for the \texttt{unitObjects} are populated by using simulations from the model.
\end{itemize}

The primary purpose of of these pre-built constructor functions is to provide source code demonstrating how users may create a variety of different \texttt{panelPomp} objects, and to enable quick testing and comparison of newly developed PanelPOMP methodology using existing models and data.
The parent \CRANpkg{pomp} package also contains a number of pre-built models and datasets for similar purposes.
In cases where there may be confusion between the constructor functions in these packages, the \texttt{panel} prefix is used to clarify the distinction.

\hypertarget{generic-methods-for-panelpomp-objects}{%
\subsection{\texorpdfstring{Generic methods for \texttt{panelPomp} objects}{Generic methods for panelPomp objects}}\label{generic-methods-for-panelpomp-objects}}

\noindent The created object from the previous section, \texttt{gomp}, is a PanelPOMP model with 3 independent units, each with a unique number of observations specified by the length of the \texttt{times} argument in their constructor function.
In this model \(\kappa_u, \tau_u\) and \(X_{u, 0}\) are treated as unit-specific, and \(r_u=r\), \(\sigma_u = \sigma\) are shared for all unit objects.
Internally, each \texttt{unit\_object} treats both types of parameters the same, and thereby the construction of the unit objects does not require renaming unit-specific parameters to allow for distinction across units.
For example, though \(\kappa_u\) may be unique for each unit \(u\), the internal functions representing the process model (Eq. \eqref{eq:proc}) can use the variable name \texttt{kappa} to represent the parameter in all unit objects, rather than needing unique variable names for each unit.
This important feature allows for changing which parameters are treated as shared and which are unit-specific without having to redefine each of the unit objects.

Model parameters can be extracted and modified using the \texttt{coef()} generic function.
The default treatment of this function is the convention \texttt{\textless{}parameter\ name\textgreater{}{[}\textless{}unit\ name\textgreater{}{]}} for unit-specific parameters, and \texttt{\textless{}parameter\ name\textgreater{}} for shared parameters.
This format is intended to closely reflect the standard mathematical notation for unit-specific parameters.
For instance, to change the value of \(\kappa_2\), we could refer to the corresponding parameter \texttt{K{[}unit2{]}} in the \texttt{gomp} object:

\begin{verbatim}
coef(gomp)['K[unit2]'] <- 0.9
coef(gomp)
\end{verbatim}

\begin{verbatim}
#>          r      sigma   K[unit1] tau[unit1] X_0[unit1]   K[unit2] tau[unit2] 
#>       0.10       0.10       1.00       0.10       1.00       0.90       0.07 
#> X_0[unit2]   K[unit3] tau[unit3] X_0[unit3] 
#>       1.00       1.20       0.15       1.00
\end{verbatim}

It can be more convenient to view unit-specific parameters as a matrix and shared parameters as a vector.
This can be done using the \texttt{format\ =\ \textquotesingle{}list\textquotesingle{}} argument of the \texttt{coef()} function, or by alternatively using the \texttt{shared()} and \texttt{specific()} functions to extract only the shared or unit-specific parameters, respectively.

\begin{verbatim}
coef(gomp, format = 'list')
\end{verbatim}

\begin{verbatim}
#> $shared
#>     r sigma 
#>   0.1   0.1 
#> 
#> $specific
#>      unit
#> param unit1 unit2 unit3
#>   K     1.0  0.90  1.20
#>   tau   0.1  0.07  0.15
#>   X_0   1.0  1.00  1.00
\end{verbatim}

\begin{verbatim}
shared(gomp)
\end{verbatim}

\begin{verbatim}
#>     r sigma 
#>   0.1   0.1
\end{verbatim}

The \texttt{shared()\textless{}-} and \texttt{specific()\textless{}-} setter functions are also convenient for modifying which model parameters are considered shared and unit-specific.

\begin{verbatim}
shared(gomp) <- c("tau" = 0.15)
shared(gomp)
\end{verbatim}

\begin{verbatim}
#>   tau     r sigma 
#>  0.15  0.10  0.10
\end{verbatim}

\begin{verbatim}
specific(gomp)
\end{verbatim}

\begin{verbatim}
#>      unit
#> param unit1 unit2 unit3
#>   K       1   0.9   1.2
#>   X_0     1   1.0   1.0
\end{verbatim}

Before altering the definitions of existing parameters, these functions verify whether the parameters are already present in the model.
They enable changing parameter values and adjusting whether a parameter is unit-specific or shared, but do not allow for the creation or deletion of model parameters.

A non-exhaustive list of useful methods that are frequently applied to \texttt{panelPomp} objects in the course of a data analysis include:

\begin{itemize}
\tightlist
\item
  \texttt{simulate()}. For \texttt{panelPomp} objects that contain units with the functions \texttt{rinit}, \texttt{rprocess}, and \texttt{rmeasure}, the \texttt{simulate()} function can generate a number of simulations, specified by the \texttt{nsim} argument, which defaults as \texttt{nsim\ =\ 1}. The resulting object is a \texttt{panelPomp} object where the simulated results are saved as data in the unit objects.
\item
  \texttt{plot()}. This function plots each unit of a \texttt{panelPomp} object sequentially. For derived classes, such as \texttt{pfilterd.ppomp} resulting from applying \texttt{pfilter} to a \texttt{panelPomp}, or \texttt{mif2d.ppomp} resulting from an application of \texttt{mif2}, diagnostic plots are produced for each unit. For example, calling \texttt{plot()} on the measles model constructed by the \texttt{panelMeasles()} function will plot the measles data available for the specified UK cities:
\end{itemize}

\begin{verbatim}
plot(panelMeasles(), units = c('Bradford', 'London'))
\end{verbatim}

\begin{center}\includegraphics{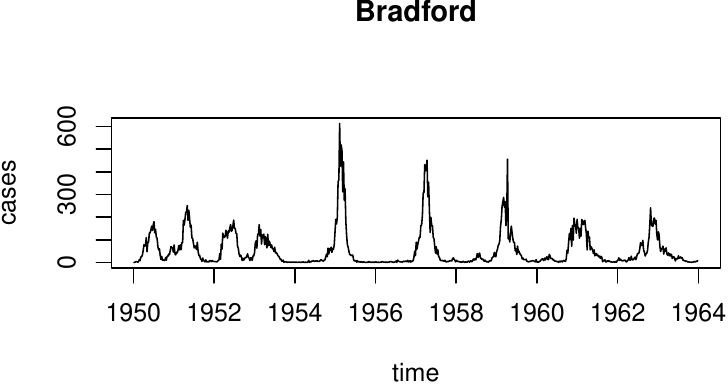} \end{center}

\begin{center}\includegraphics{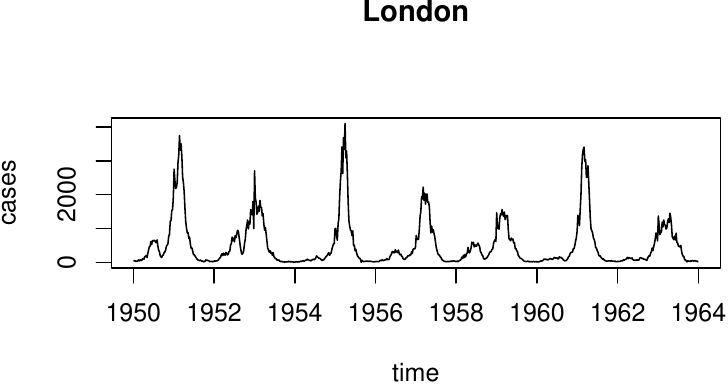} \end{center}

\begin{itemize}
\tightlist
\item
  \texttt{as()}. Many essential features of the basic model components of a \texttt{panelPomp} object have already undergone extensive development and testing in the context of the \CRANpkg{pomp} package.
  To avoid unnecessary duplication, the \texttt{as()} function offers a way to convert a \texttt{panelPomp} object into a list of \texttt{pomp} objects.
  For example, \texttt{as(gomp,\ "list")} and \texttt{as(gomp,\ "pompList")} will convert the \texttt{gomp} \texttt{panelPomp} object into a \texttt{list} or \texttt{pompList}, respectively.
  This is particularly useful when it is appropriate to consider the individual components of the \texttt{panelPomp} object.
\end{itemize}

\hypertarget{inference-methodology}{%
\section{Inference methodology}\label{inference-methodology}}

All POMP methods can in principle be extended to PanelPOMPs;
three different ways to represent a PanelPOMP as a POMP were identified by \citet{romero-severson15}.
The ability to represent a PanelPOMP models as a POMP model, however, does not imply that methodology for POMP models will be feasible for PanelPOMP models.
In particular, sequential Monte Carlo algorithms can have prohibitive scaling difficulties with the high dimensionality that arises in PanelPOMP models.

The current version of \CRANpkg{panelPomp} emphasizes plug-and-play methods \citep{breto09, he10}, also known as likelihood-free \citep{marjoram03, sisson07}, that are applicable to dynamic models for which a simulator is available even when the transition densities are unavailable.
In the terminology used in this article, this means that \texttt{dprocess} (Eq. \eqref{eq:proc}) is not needed in order to perform inference.
This class of algorithms includes methods such as a particle filter \citep{arulampalam02} and the panel iterated filter (PIF) \citep{breto20}, which are methods that can be used to evaluate and maximize model likelihoods, respectively.
In this section, we describe and demonstrate a plug-and-play likelihood-based inference workflow using the Stochastic Gompertz population model described previously.
To highlight the capability of \CRANpkg{panelPomp} to perform inference on high-dimensional models, we reconstruct a model with \(U = 50\) measurement units and \(N = N_u = 100\) observations per unit using the \texttt{panelGompertz()} constructor function.

\begin{verbatim}
gomp <- panelGompertz(N = 100, U = 50)
\end{verbatim}

\hypertarget{log-likelihood-evaluation-via-particle-filters}{%
\subsection{Log-likelihood evaluation via particle filters}\label{log-likelihood-evaluation-via-particle-filters}}

The particle filter, also known as sequential Monte Carlo, is a standard tool for log-likelihood evaluation on nonlinear non-Gaussian POMP models.
The log-likelihood function is a central component of Bayesian and frequentist inference.
Due to the dynamic independence between units that is a defining feature of a PanelPOMP model, particle filtering can be carried out separately on each unit.
The \texttt{pfilter} method for \texttt{panelPomp} objects is therefore a direct extension of the \texttt{pfilter} method for \texttt{pomp} objects from the \texttt{pomp} package.
All mathematical functions needed to carry out the particle filter for a PanelPOMP model are represented by the functions \texttt{rinit}, \texttt{rprocess}, \texttt{rmeasure} and \texttt{dmeasure} (Eqs. \eqref{eq:proc}--\eqref{eq:init}).

Methods applied to \texttt{panelPomp} objects---like the \texttt{pfilter} function---typically create new objects of the same class or a child class of the original object.
For instance, we can perform a particle filter on the \texttt{gomp} object that contains 50 units, and 100 observations per unit in the following way:

\begin{verbatim}
gomp_pfd <- pfilter(gomp, Np = 1000)
\end{verbatim}

In this case, we used \texttt{Np} = 1000 particles to obtain a single stochastic estimate of the log-likelihood of the \texttt{gomp} model.
The resulting object \texttt{gomp\_pfd} is of class \texttt{pfilterd.ppomp}, which is a child class of \texttt{panelPomp} and contains the estimated log-likelihood of each unit object, as well as the log-likelihood of the entire panel;
these can be accessed using the \texttt{unitLogLik()} and \texttt{logLik()} functions, respectively.

In practice it is advisable to repeat this Monte Carlo approximation in order to reduce and quantify the error associated with the estimate.
Because sequential Monte Carlo algorithms can be computationally expensive, we obtain replicated estimates by taking advantage of multicore computation using the \CRANpkg{foreach} \citep{foreach22} and \CRANpkg{doParallel} \citep{dopar22} packages:

\begin{verbatim}
pf_results <- foreach(i = 1:10) %dopar% {
    pfilter(gomp, Np = 1000)
}
\end{verbatim}

\noindent This took 2.68 seconds to run the 10 replicates in parallel, resulting in a list of objects of class \texttt{pfilterd.ppomp}.
We can use the \texttt{logLik} function to extract the Monte Carlo estimate of the log-likelihood \(\lambda^{[i]}\) for each replicate \(i\), and \texttt{unitLogLik} to extract the vector of component Monte Carlo log-likelihood estimates \(\lambda^{[i]}_{u}\) for each unit \(u=1,\dots,U\), where \(\lambda^{[i]}=\sum_{u=1}^{U}\lambda^{[i]}_{u}\).
For a POMP model, replicated log-likelihood evaluations via the particle filter are usually averaged on the natural scale, rather than the log scale, to take advantage of the unbiasedness of the particle filter likelihood estimate.
Thus, we have

\[
\hat{\lambda}_1 = \log \Big(\frac{1}{I}\sum_{i=1}^{I}
\exp
\big\{
  \sum_{u=1}^{U} \lambda^{[i]}_{u}
\big\}\Big),
\]

\noindent which can be implemented as

\begin{verbatim}
lambda_1 <- logmeanexp(
  sapply(pf_results,logLik), se = TRUE
)
\end{verbatim}

\noindent giving \(\hat{\lambda}_1=2065.4\) with a jack-knife standard error of 0.5.
Taking advantage of the independence of the units in the panel structure, \citet{breto20} showed it is preferable to average the replicates of marginal likelihood for each unit before taking a product over units.
This corresponds to
\[
\hat{\lambda}_2 = \log \Big(\prod_{u=1}^U \frac{1}{I}\sum_{i=1}^{I}  \exp \big\{ \hat{\lambda}^{[i]}_{u} \big\}\Big),
\]
which can be obtained using

\begin{verbatim}
lambda_2 <- panel_logmeanexp(
  sapply(pf_results,unitLogLik), MARGIN = 1, se = TRUE
)
\end{verbatim}

\noindent giving \(\hat{\lambda}_2=2068.5\) with a jack-knife standard error of 1.1.
For this model, a Kalman filter log-likelihood evaluation gives an exact answer, \(\lambda=2068.2\).

\hypertarget{maximum-likelihood-estimation}{%
\subsection{Maximum likelihood estimation}\label{maximum-likelihood-estimation}}

Although particle filters can effectively approximate the log-likelihood of non-linear models, it is well known that obtaining a maximum likelihood estimate for fixed parameters using these filters is difficult in practice.
To maximize the likelihood, we employ iterated filtering algorithms, which perform repeated particle filtering on an extended version of the model that incorporates time-varying parameter perturbations.
With each iteration, the magnitude of these perturbations is reduced, allowing the algorithm to approach a local maximum of the likelihood function.
An example of this type of algorithm for general POMP models includes the IF2 iterated filtering algorithm \citep{ionides15}.
IF2 has been successfully employed for likelihood-based inference in various POMP models, particularly in epidemiology and ecology, as reviewed by \citet{breto18}.
However, both particle filters and IF2 face scalability issues as model dimensions increase and IF2 cannot be applied to each unit separately when the PanelPOMP model includes shared parameter values.

A panel iterated filtering (PIF) algorithm was developed by \citet{breto20}, extending IF2 to panel data.
An implementation of PIF in \CRANpkg{panelPomp} is provided by the \texttt{mif2} method for class \texttt{panelPomp}, following the pseudocode in Algorithm \ref{alg:pif}.
The pseudocode sometimes omits explicit specification of ranges over which variables are to be computed when this is apparent from the context:
it is understood that \(j\) takes values in \(1\,{:}\,J\), \(a\) in \(1\,{:}\,A\) and \(b\) in \(1\,{:}\,B\).
The \(N[0,1]\) notation corresponds to the construction of independent standard normal random variables, leading to Gaussian perturbations of parameters on a transformed scale.
The theory allows considerable flexibility in how the parameters are perturbed, but Gaussian perturbations on an appropriate scale are typically adequate.

\begin{algorithm}[t!]
  \caption{
    \texttt{mif2$\big($pp, Nmif\,=\,$M$, Np\,=\,$J$,
start{\,=\,}$(\phi^0_{a},\psi^0_{b,u})$,
rw\_sd{\,=\,}$(\sigma^\Phi_{a,n},\sigma^\Psi_{b,u,n})$,
cooling.factor.50{\,=\,}$\rho^{50}\big)$}, where \texttt{pp} is a \texttt{panelPomp} object containing data and defined \texttt{rprocess}, \texttt{dmeasure}, \texttt{rinit} and \texttt{partrans} components.
    \label{alg:pif}
    }
\noindent\begin{tabular}{ll}
{\bf input:}\rule[-1.5mm]{0mm}{6mm}
& Data, $y_{u,n}^*$, $u$ in $1\,{:}\,U$, $n$ in $1\,{:}\,N$\\
& Simulator of initial density, $f_{X_{u,0}}(x_{u,0} {\,;\,} \phi,\psi_{u})$ \\
& Simulator of transition density, $f_{X_{u,n}|X_{u,n-1}}(x_{u,n}\,|\, x_{u,n-1}{\,;\,} \phi,\psi_{u})$ \\
& Evaluator of measurement density, $f_{Y_{u,n}|X_{u,n}}(y_{u,n}\,|\, x_{u,n}{\,;\,}\phi,\psi_{u})$ \\
& Number of particles, $J$, and number of iterations, $M$\\
& Starting shared parameter swarm, $\Phi^0_{a,j}=\phi^0_{a}$, $a$ in $1\,{:}\,A$, $j$ in $1\,{:}\,J$\\
& Starting unit-specific parameter swarm, $\Psi^0_{b,u,j}=\psi^0_{b,u}$,  $b$ in  $1\,{:}\,B$, $j$ in $1\,{:}\,J$\\
& Random walk intensities,
$\sigma^\Phi_{a,n}$ and $\sigma^\Psi_{b,u,n}$ \\
& Parameter transformations, $h^{\Phi}_{a}$ and $h^{\Psi}_{b}$, with inverses
 $\big(h^{\Phi}_{a}\big)^{-1}$ and $\big(h^{\Psi}_{b}\big)^{-1}$
\\
&Logical variable determining marginalization, MARGINALIZE
\\
{\bf output:}\rule[-1.5mm]{0mm}{6mm}
& Final parameter swarm, $\Phi^M_{a,j}\;$ and $\;\Psi^M_{b,u,j}$
\rule[-2mm]{0mm}{5mm}
\end{tabular}
\hrule

\noindent\begin{tabular}{l}
For $m$ in $1: M$\rule[0mm]{0mm}{5mm}\\
\hspace{6mm} $\Phi^m_{a,0,j}=\Phi^{m-1}_{a,j}$
\rule[-2mm]{0mm}{5mm} \\
\hspace{6mm} For $u$ in $1:U$\\
\hspace{6mm} \hspace{6mm} $\Phi^{F,m}_{a,u,0,j} =\big(h^{\Phi}_{a}\big)^{-1}
  \left(
  h^\Phi_a \big(\Phi^{m}_{a,u-1,j}\big)+
    \rho^{m}\sigma^{\Phi}_{a,0} Z^{\Phi,m}_{a,u,0,j}
  \right)$ for $ Z^{\Phi,m}_{a,u,0,j}\sim N[0,1]$ \rule[-3mm]{0mm}{5mm} \\
\hspace{6mm} \hspace{6mm} $\Psi^{F,m}_{b,u,0,j} =\big(h^{\Psi}_{b}\big)^{-1}
  \left(
  h^\Psi_b \big(\Psi^{m-1}_{b,u,j}\big)+
    \rho^{m}\sigma^{\Psi}_{b,u,0} Z^{\Psi,m}_{b,u,0,j}
  \right)$ for $ Z^{\Psi,m}_{b,u,0,j}\sim N[0,1]$ \rule[-3mm]{0mm}{5mm} \\
\hspace{6mm} \hspace{6mm} $X_{u,0,j}^{F,m}\sim f_{X_{u,0}}
  \left(
    x_{u,0} \; \big| \; \Phi^{F,m}_{a,u,0,j} \, ,
    \Psi^{F,m}_{b,u,0,j}
  \right)$ \rule[-3mm]{0mm}{5mm} \\
\hspace{6mm} \hspace{6mm} For $n$ in $1: N_{u}$\\
\hspace{6mm} \hspace{6mm} \hspace{6mm} $\Phi^{P,m}_{a,u,n,j} =\big(h^{\Phi}_{a}\big)^{-1}
  \left(
  h^\Phi_a \big(\Phi^{F,m}_{a,u,n-1,j}\big)+
    \rho^{m}\sigma^{\Phi}_{a,n} Z^{\Phi,m}_{a,u,n,j}
  \right)$ for $ Z^{\Phi,m}_{a,u,n,j}\sim N[0,1]$ \rule[-3mm]{0mm}{5mm} \\
\hspace{6mm} \hspace{6mm} \hspace{6mm} $\Psi^{P,m}_{b,u,n,j} =\big(h^{\Psi}_{b}\big)^{-1}
  \left(
  h^\Psi_b \big(\Psi^{F,m}_{b,u,n-1,j}\big)+
    \rho^{m}\sigma^{\Psi}_{b,u,n} Z^{\Psi,m}_{b,u,n,j}
  \right)$ for $ Z^{\Psi,m}_{b,u,n,j}\sim N[0,1]$ \rule[-3mm]{0mm}{5mm} \\
\hspace{6mm} \hspace{6mm} \hspace{6mm} $X_{u,n,j}^{P,m}\sim f_{X_{u,n}|X_{u,n-1}}
   \left(
     x_{u,n} \; \big| \; X^{F,m}_{u,n-1,j} \; {\,;\,} \;
     \Phi^{P,m}_{a,u,n,j} \, ,
     \Psi^{P,m}_{b,u,n,j}
   \right)$  \rule[-3mm]{0mm}{5mm} \\
\hspace{6mm} \hspace{6mm} \hspace{6mm} $w_{u,n,j}^m = f_{Y_{u,n}|X_{u,n}}
    \left(y^*_{u,n} \; \big| \; X_{u,n,j}^{P,m} \; {\,;\,} \;
       \Phi^{P,m}_{a,u,n,j} \, ,
       \Psi^{P,m}_{b,u,n,j}
    \right)$  \rule[-3mm]{0mm}{5mm} \\
\hspace{6mm} \hspace{6mm} \hspace{6mm} Draw $k_{1:J}$ with $\mathbb{P}(k_j=i)=  w_{u,n,i}^m\Big/\sum_{q=1}^J w_{u,n,q}^m$ \rule[-3mm]{0mm}{5mm}  \\
\hspace{6mm} \hspace{6mm} \hspace{6mm} $\Phi^{F,m}_{a,u,n,j}=\Phi^{P,m}_{a,u,n,k_{j}}$\,,  $\;\; \Psi^{F,m}_{b,u,n,j}=\Psi^{P,m}_{b,u,n,k_{j}} \;$ and $\; X^{F,m}_{u,n,j}=X^{P,m}_{u,n,k_j}$ \rule[-3mm]{0mm}{5mm}  \\
\hspace{6mm} \hspace{6mm} \hspace{6mm} If MARGINALIZE then\\
\hspace{6mm} \hspace{6mm} \hspace{6mm} \hspace{6mm} $\Psi^{F,m}_{b,\nu,n,j}=\Psi^{P,m}_{b,\nu,n,j}$ for all $\nu \neq u$\\
\hspace{6mm} \hspace{6mm} \hspace{6mm} Else\\
\hspace{6mm} \hspace{6mm} \hspace{6mm} \hspace{6mm} $\Psi^{F,m}_{b,\nu,n,j}=\Psi^{P,m}_{b,\nu,n,k_j}$ for all $\nu \neq u$\\
\hspace{6mm} \hspace{6mm} End For \\ 
\hspace{6mm} \hspace{6mm}  $\Phi^{m}_{a,u,j}=\Phi^{F,m}_{a,u,N_{u},j}$ and
           $\Psi^{m}_{b,u,j}=\Psi^{F,m}_{b,u,N_{u},j}$
 \\
\hspace{6mm} End For \\ 
\hspace{6mm} $\Phi^{m}_{a,j}=\Phi^m_{a,U,j}$
\\
End For \\ 
\end{tabular}
\end{algorithm}

At a conceptual level, the PIF algorithm has an evolutionary analogy: successive iterations mutate parameters and select among the fittest outcomes measured by Monte Carlo likelihood evaluation.
Most often, the perturbation parameters \(\sigma^\Phi_{a,n}\) and \(\sigma^\Psi_{b,u,n}\) in Algorithm \ref{alg:pif} will not depend on \(n\).
For parameters that have uncertainty on a unit scale, the value 0.02 demonstrated here has been commonly used.
The help documentation on the \texttt{rw\_sd} argument gives instruction on using additional structure should it become necessary.

The unmarginalized PIF was proposed by \citet{breto20}, who provided theoretical convergence results for the algorithm.
When MARGINALIZE = TRUE, the PIF algorithm is modified such that the particles representing the unit-specific parameter \(\psi_{b, u}\) remain unchanged when filtering through a unit \(\nu \neq u\).
Recent results suggest the marginalized PIF (MPIF) algorithm has superior empirical performance in many situations, but does not yet have theoretical support.
For the remainder of this article, the presented results use the unmarginalized version of the algorithm.

Parameter transformations (\(h^{\Phi}_{a}\) and \(h^{\Psi}_{b}\)) are used to ensure that parameter estimates remain within accepted bounds.
For example, parameters that are required to be positive can be estimated on a log-transformed scale to maintain their positivity when converted back to the natural scale.
This process is facilitated by the \texttt{parameter\_trans} function, which manages the transformation automatically.
This relieves users from the need to handle parameters on the transformed scale directly; they only need to specify the desired transformation.
In the case of the \texttt{gomp} model, all model parameters must be non-negative.
This requirement can be met by creating unit objects and assigning the \texttt{partrans} argument as follows:

\begin{verbatim}
parameter_trans(log = c("K", "r", "sigma", "tau", "X.0"))
\end{verbatim}

This setup ensures that the parameters \(\kappa_u\), \(r_u\), \(\sigma_u\), \(\tau_u\), and \(X_{u,0}\) are all estimated on a log-transformed scale, thereby maintaining their non-negativity on the natural scale.
Other common parameter transformations that can be implemented include the \texttt{logit} transformation, which ensures parameters are in the interval \((0, 1)\), and the \texttt{barycentric} transformation, which is used when a collection of parameters must lie in the interval \((0, 1)\), with the additional constraint that they sum to one.
Finally, custom transformations can be defined using the \texttt{toEst} and \texttt{fromEst} arguments to the \texttt{parameter\_trans} function.

We demonstrate maximum likelihood estimation using parameter transformations for the \texttt{gomp} object described previously.
To do so, we need to specify starting parameter values from where to start the search.
This can be done by explicitly providing the starting parameters using either the \texttt{start} or the \texttt{shared.start} and \texttt{specific.start} arguments of the \texttt{mif2()} function.
Alternatively, the existing parameter values in the \texttt{panelPomp} object will be implicitly used as a starting point, as done in the following example.
For simplicity, we fix \(\kappa_{u}=1\) and the initial condition \(X_{u,0}=1\), maximizing over two shared parameters, \(r\) and \(\sigma\), and one unit-specific parameter \(\tau_u\), starting from their default values in the \texttt{gomp} object:

\begin{verbatim}
gomp_mif2d <- mif2(
  gomp,  # panelPomp model, which contains parameter transformations and values.
  Nmif = 25,  # Number of Iterations (M)
  Np = 250,  # Number of Particles (J)
  cooling.fraction.50 = 0.5,  # Cooling intensity, after 50 iterations
  cooling.type = "geometric",  # Cooling Style
  rw.sd = rw_sd(r = 0.02, sigma = 0.02, tau = 0.02)  # Random Walk SD
)
\end{verbatim}

\noindent The output \texttt{gomp\_mif2d} is a \texttt{mif2d.ppomp} object, which is a child class of \texttt{panelPomp}.
The algorithmic parameters are very similar to those of the \texttt{mif2} method for class \texttt{pomp}.
The perturbations, determined by the \texttt{rw\_sd} argument, may be a list giving separate instructions for each unit.
When only one specification for a unit-specific parameter is given (as we do for \(\tau_u\) here) the same perturbation is used for all units.
As such, functions for coefficient extraction can be used to get the final parameter estimates, for instance:

\begin{verbatim}
shared(gomp_mif2d)
\end{verbatim}

\begin{verbatim}
#>          r      sigma 
#> 0.06833055 0.09123509
\end{verbatim}

For Monte Carlo maximization, replication from diverse starting points is recommended.
Further, larger values of \texttt{Nmif} and \texttt{Np} are typically needed in order to reliably maximize model likelihoods in practice.
Smaller initial searches for parameter estimates are useful in that they can be used to estimate the computational cost of a larger search or to help determine values of hyperparameters.
The computational complexity of the PIF algorithm is \(O(JMNU)\), and so the cost of a larger search may be estimated by noting that the complexity is linear in each of the arguments \(J\) and \(M\).

We demonstrate such a maximization search on \texttt{gomp}.
The small, single PIF maximization took 40.8 seconds to compute.
By increasing the number of iterations (\texttt{Nmif}) and particles (\texttt{Np}) each by a factor of \(6\), we expect the computation time for a single parameter initialization to take roughly 24.5 minutes, noting that there is some overhead in the maximization function that is not increased on larger jobs.
Using 36 cores, we then expect the maximization routine to take 24.5 minutes for 36 distinct starting values.

To define diverse starting points for the Monte Carlo replicates, we make uniform draws from a specified box.
The parameter bounds on this box are not intended to be bounds on the final parameter estimates, as there is a possibility that the data lead the parameter search elsewhere.
However, if replicated searches started from this box reliably reach a consensus, we claim we have carefully investigated this part of parameter space.
A larger box leads to greater confidence that the relevant part of the parameter space has been searched, at the expense of requiring additional work.
The \texttt{runif\_panel\_design()} function facilitates the construction and drawing random points from within the box.

\begin{verbatim}
starts <- runif_panel_design(
  lower = c('r' = 0.05, 'sigma' = 0.05, 'tau' = 0.05, 'K' = 1, 'X.0' = 1),
  upper = c('r' =  0.2, 'sigma' =  0.2, 'tau' =  0.2, 'K' = 1, 'X.0' = 1),
  specific_names = c('K', 'tau', 'X.0'),
  unit_names = names(gomp),
  nseq = 36
)
\end{verbatim}

\noindent We then carry out a search from each starting point:

\begin{verbatim}
mif_results <- foreach(start=iter(starts,"row")) %dopar% {
  mif2(
    gomp, start = unlist(start),
    Nmif = 150,
    Np = 1500,
    cooling.fraction.50 = 0.5,
    cooling.type = "geometric",
    transform = TRUE,
    rw.sd = rw_sd(r = 0.02, sigma = 0.02, tau = 0.02)
  )
}
\end{verbatim}

\noindent This took 15.6 minutes using 36 cores, producing a list of objects of class \texttt{mifd.ppomp}.
We can check on convergence of the searches, and possibly diagnose improvements in the choices of algorithmic parameters, by consulting trace plots of the searches available via the \texttt{traces} method for class \texttt{mifd.ppomp}.
This follows recommendations by \citet{ionides06} and \citet{king16}.

\hypertarget{block-optimization}{%
\subsubsection{Block optimization}\label{block-optimization}}

A characteristic of PanelPOMP models is the large number of parameters arising when unit-specific parameters are specified for a large number of units.
For a fixed value of the shared parameters, the likelihood of the unit-specific parameters factorizes over the units.
The factorized likelihood can be maximized separately over each unit, replacing a challenging high-dimensional problem with many relatively routine low-dimensional problems.
This suggests a block maximization strategy where unit-specific parameters for each unit are maximized as a block.
\citet{breto20} used a simple block strategy where a global search over all parameters is followed by a block maximization over units for unit-specific parameters.
The theoretical guarantees available for the PIF algorithm \citep{breto20} imply that this block-refinement strategy is not necessary for convergence, but empirically it can help reduce the computational effort needed to fully maximize the likelihood.

We demonstrate this here, refining each of the maximization replicates above.
The following function carries out a maximization search of unit-specific parameters for a single unit.
The call to \texttt{mif2} takes advantage of argument recycling: all algorithmic parameters are re-used from the construction of \texttt{mifd\_gomp} except for the re-specified random walk standard deviations which ensures that only the unit-specific parameters are perturbed.

\begin{verbatim}
mif_unit <- function(unit, mifd_gomp, reps = 6) {
  unit_gomp <- unit_objects(mifd_gomp)[[unit]]
  
  mifs <- replicate(
    n = reps, mif2(unit_gomp, rw.sd = rw_sd(tau = 0.02))
  )
  
  best <- which.max(sapply(mifs, logLik))
  coef(mifs[[best]])["tau"]
}
\end{verbatim}

Now we apply this block maximization to find updated unit-specific parameters for each replicate, and we insert these back into the \texttt{panelPomp}.

\begin{verbatim}
mif_block <- foreach(mf=mif_results) %dopar% {
  mf@specific["tau",] <- sapply(1:length(mf), mif_unit, mifd_gomp = mf)
  mf
}
\end{verbatim}

\noindent This took 40.4 seconds to compute all 36 block refinements in parallel.

We expect Monte Carlo estimates of the maximized log-likelihood functions to fall below the actual (usually unknown) value.
This is in part because imperfect maximization can only reduce the maximized likelihood, and in part a consequence of Jensen's inequality applied to the likelihood evaluation: the unbiased SMC likelihood evaluation has a negative bias on estimation of the log-likelihood.

\hypertarget{parameter-uncertainty}{%
\subsection{Parameter uncertainty}\label{parameter-uncertainty}}

A key component of a likelihood-based inference framework is the estimation of parameter uncertainty, often achieved by the estimation of confidence intervals.
This task is particularly challenging for general non-linear, latent variable models.
Some potential methods include profile likelihoods, observed Fisher information, and the bootstrap method.
Here, we demonstrate the profile likelihood approach, which has proven useful for mechanistic models \citep{simpson23}.

The profile likelihood function is constructed by fixing a focal parameter at various values and then maximizing the likelihood over all other parameters for each value of the focal parameter.
Constructing a profile likelihood function offers several practical advantages:

\begin{itemize}
\tightlist
\item
  Evaluations at neighboring values of the focal parameter provide additional Monte Carlo replication. Typically, the true profile log-likelihood is smooth, and asymptotically close to quadratic under regularity conditions, so deviations from a smooth fitted line can be interpreted as Monte Carlo error.
\item
  Large-scale features of the profile likelihood reveal a region of the parameter space outside which the model provides a poor explanation of the data.
\item
  Co-plots, which show how the values of other maximized parameters vary along the profile, may provide insights into parameter trade-offs implied by the data.
\item
  The smoothed Monte Carlo profile log-likelihood can be used to construct an approximate 95\% confidence interval. The resulting confidence interval can be properly adjusted to accommodate both statistical and Monte Carlo uncertainty \citep{ionides17}.
\end{itemize}

Once we have code for maximizing the likelihood, only minor adaptation is needed to carry out the maximizations for a profile.
The \texttt{runif\_panel\_design} generating the starting values is replaced by a call to \texttt{profile\_design}, which assigns the focal parameter to a grid of values and randomizes the remaining parameters.
The random walk standard deviation for the focal parameter is unassigned, which leads it to be set to zero and therefore the parameter remains fixed during the maximization process.
The following code combines the joint and block maximizations developed above.

\begin{verbatim}
# Names of the estimated parameters
estimated <- c(
  "r", "sigma", paste0("tau[unit", 1:length(gomp), "]")
)

# Names of the fixed parameters (not estimated)
fixed <- names(coef(gomp))[!names(coef(gomp)) %in% estimated]

profile_starts <- profile_design(
  r = seq(0.05, 0.2, length = 20),
  lower = c(coef(gomp)[estimated] / 2, coef(gomp)[fixed])[-1],
  upper = c(coef(gomp)[estimated] * 2, coef(gomp)[fixed])[-1],
  nprof = 5, type = "runif"
)

profile_results <- foreach(start = iter(profile_starts, "row")) %dopar% {
  mf <- mif2(
    mif_results[[1]],
    start = unlist(start),
    rw.sd = rw_sd(sigma = 0.02, tau = 0.02)
  )
  mf@specific["tau", ] <- sapply(1:length(mf), mif_unit, mifd_gomp = mf)
  mf
}
\end{verbatim}

In the above code, the \texttt{profile\_design} function creates \(20 \times 5 = 100\) unique starting points to perform the profile search.
Parallelized over 36 cores, these \(100\) searches took 47.6 total minutes.
However, we are not quite done gathering the results for the profile.
The perturbed filtering carried out by \texttt{mif2} leads to an approximate likelihood evaluation, but additional accuracy is obtained by re-evaluating the likelihood without perturbations.
Also, replication is recommended to reduce and quantify Monte Carlo error.
We do this, and tabulate the results.

\begin{verbatim}
profile_table <- foreach(mf=profile_results,.combine=rbind) %dopar% {
  LL <- replicate(10, logLik(pfilter(mf, Np = 2500)))
  LL <- logmeanexp(LL, se = TRUE)
  data.frame(t(coef(mf)), loglik = LL[1], loglik.se = LL[2])
}
\end{verbatim}

The likelihood evaluations took 2.5 minutes.
It is appropriate to spend comparable time evaluating the likelihood to the time spent maximizing it: a high quality maximization without high quality likelihood evaluation is hard to interpret, whereas good evaluations of the likelihood in a vicinity of the maximum can inform about the shape of the likelihood surface in this region, and this may be as relevant as knowing the exact maximum.

The Monte Carlo adjusted profile (MCAP) approach of \citet{ionides17} is implemented by the \texttt{mcap()} function in \CRANpkg{pomp}.
This function constructs a smoothed profile likelihood, by application of the \texttt{loess} smoother.
It computes a local quadratic approximation that is used to derive an extension to the classical profile likelihood confidence interval that makes allowance for Monte Carlo error in the calculation of the profile points.
Theoretically, an MCAP procedure can obtain statistically efficient confidence intervals even when the Monte Carlo error in the profile likelihood is asymptotically growing and unbounded \citep{ning21}.
Log-likelihood evaluation has negative bias, as a consequence of Jensen's inequality for an unbiased likelihood estimate.
This bias produces a vertical shift in the estimated profile, which fortunately does not have consequence for the confidence interval if the bias is slowly varying.

\begin{figure}

{\centering \includegraphics{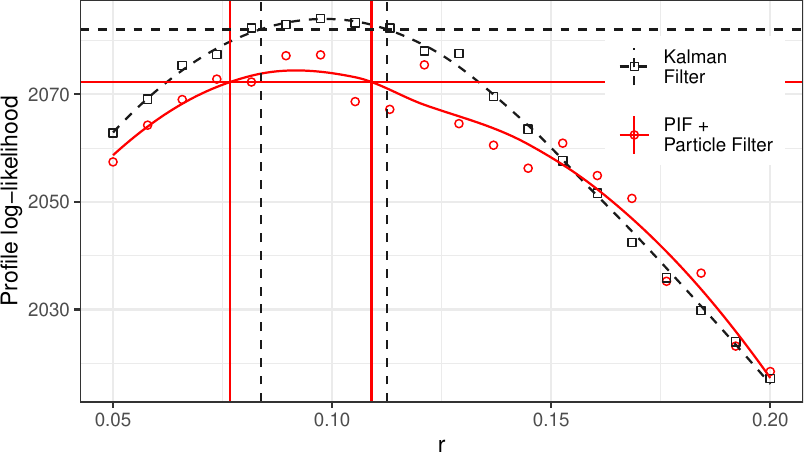} 

}

\caption{The Monte Carlo adjusted profile confidence interval (solid red lines, evaluation points shown as circles). Construction using deterministic optimization of the likelihood calculated by the Kalman filter (dashed lines, evaluation points show as squares).}\label{fig:mcap-plot}
\end{figure}

The profile points evaluated above, and stored in \texttt{profile\_table}, can be used to compute a 95\% MCAP confidence interval as follows:

\begin{verbatim}
profile_table <- profile_table |> 
  as.data.frame() |>
  dplyr::group_by(r) |> 
  dplyr::slice_max(n = 1, order_by = loglik)

gomp_mcap <- pomp::mcap(
  logLik = profile_table$loglik,
  parameter = profile_table$r,
  level = 0.95
)
\end{verbatim}

\noindent The construction of the confidence interval is best shown by a plot of the smoothed profile likelihood, shown in Fig. \ref{fig:mcap-plot} \citep{wickham16}.
In this toy example, the exact likelihood can be calculated using the Kalman filter, and this is carried out by the \texttt{panelGompertzLikelihood} function.
The likelihood can then be maximized using a general-purpose optimization procedure such as \texttt{optim()} in R.
With large numbers of parameters, and no guarantee of convexity, this numerical optimization is not entirely routine.
One might consider a block optimization strategy, but here we carry out a simple global search, which took 1.7 minutes to compute the profile likelihood, once parallelized.
The deterministic search is also not entirely smooth, and so we apply MCAP as for the Monte Carlo search.
Both deterministic and Monte Carlo optimizations can benefit from a block optimization strategy which alternates between shared and unit-specific parameters \citep{breto20}.
Such algorithms can be built using the \CRANpkg{panelPomp} functions we have demonstrated, and they will be incorporated into the package once they have been more extensively researched.

\hypertarget{conclusion}{%
\section{Conclusion}\label{conclusion}}

The analysis using the \texttt{gomp} model illustrates one approach to plug-and-play inference for PanelPOMP models, but the scope of \CRANpkg{panelPomp} is far from limited to this approach.
\CRANpkg{panelPomp} is a general and extensible framework which encourages the development of additional functionality.
The \texttt{panelPomp} class, along with its associated workhorse functions, provide an adaptable interface that can accommodate future methodologies.
In this sense, \CRANpkg{panelPomp} provides an environment for sharing and developing PanelPOMP models and methods, both through future contributions to the \CRANpkg{panelPomp} package and through open-source applications that leverage the package.
This framework will facilitate the comparison of new methodologies with existing ones, promoting continuous improvement and innovation.
Other software packages, such as \CRANpkg{pomp}, \CRANpkg{spatPomp}, \CRANpkg{nimble}, among others summarized in Section 4.2 of \citet{newman23}, can also be used to perform inference on non-linear mechanistic models.
However, none of these packages specifically address the unique challenges posed by high-dimensional PanelPOMP models.

A special class of POMP models arises when the latent process takes values in a discrete and finite space.
A model of this type is often referred to as a hidden Markov model (HMM) \citep{eddy04, doucet01, glennie23, newman23}, though this terminology has also been used as a synonym with POMP \citep{king16}.
Under this additional constraint, efficient dynamic-programming algorithms can be used to perform inference, as summarized by \citet{mcclintock20}.
In principle, \CRANpkg{panelPomp} can be leveraged to implement these existing approaches for longitudinal data, though the current version of the package emphasizes methodologies for models that have latent process with states taking values in spaces that cannot be programmatically searched.

In our example, likelihood evaluation and maximization was used to construct confidence intervals.
These calculations also provide a foundation for other techniques of likelihood-based inference, such as likelihood ratio hypothesis tests and model selection via Akaike's information criterion (AIC).
The examples discussed provide case studies in the use of these methods for scientific work.

Data analysis using large data sets or complex models may require considerable computing time.
Simulation-based methodology is necessarily computationally intensive, and access to a cluster computing environment extends the size of problems that can be tackled.
Our example workflow has a simple parallel structure that can readily take advantage of additional resources.
Embarrassingly parallel computations, such as computing the profile likelihood function at a grid of points, or replicated evaluations of the likelihood function, can be parallelized using the \CRANpkg{foreach} package.

Panel data is widely available: for many experimental and observational systems it is more practical to collect short time series on many units than to obtain one long time series.
For time series data, fitting mechanistic models specified as partially observed Markov processes has found numerous applications for formulating and answering scientific hypotheses \citep{breto09, king16}.
However, there are remarkably few examples in the literature fitting mechanistic nonlinear non-Gaussian partially observed stochastic dynamic models to panel data.
The \CRANpkg{panelPomp} package offers opportunities to remedy this situation.

\hypertarget{availability-documentation-and-code-quality-control}{%
\section{Availability, documentation and code quality control}\label{availability-documentation-and-code-quality-control}}

\CRANpkg{panelPomp} is available on CRAN and can be installed by executing \texttt{install.packages(\textquotesingle{}panelPomp\textquotesingle{})}.
The source code and developmental version of the package are available on GitHub: \url{https://github.com/panelPomp-org/panelPomp}.
Package documentation is created using \texttt{roxygen2} and is shipped with the installation of the package, but can also be found at the package website: \url{https://panelpomp-org.github.io}.
Two tutorials are provided on the website; an elementary ``Getting Started'' guide, and an in-depth introduction which contains examples that are similar to those in this article \citep{breto24}.
Continuous integration based on GitHub actions is used to build and test the package.
As of writing, unit tests have a 100\% line coverage (measured by the \CRANpkg{covr} package).

All major computations were performed on a high-performance computing (HPC) node equipped with 2x 3.0 GHz Intel Xeon Gold 6154 processors, totaling 36 cores. The system ran on Red Hat Enterprise Linux 8.8 (Ootpa) on an x86\_64-pc-linux-gnu platform, with 180 GB RAM.
We used R version 4.4.0 (2024-04-24) for our analyses.
This paper introduces \CRANpkg{panelPomp} version \texttt{1.7.0.0}, and requires \CRANpkg{pomp} version 6.2.1.0 or higher.

\hypertarget{acknowledgements-and-funding}{%
\section{Acknowledgements and funding}\label{acknowledgements-and-funding}}

This work was supported by National Science Foundation grants DMS-1761603 and DMS-1646108; National Institutes of Health grants 1-U54-GM111274, 1-U01-GM110712, and 1-R01-AI143852; and by MCIN/AEI/10.13039/501100011033 grants PID2020-116242RB-I00 and PID2023-152348NB-I00.

This research was additionally supported in part through computational resources and services provided by Advanced Research Computing at the University of Michigan, Ann Arbor.

\bibliography{ms.bib}

\address{%
Carles Bretó\\
Universitat de València\\%
Department of Economic Analysis\\ Valencia, Spain\\
\textit{ORCiD: \href{https://orcid.org/0000-0003-4695-4902}{0000-0003-4695-4902}}\\%
\href{mailto:carles.breto@uv.es}{\nolinkurl{carles.breto@uv.es}}%
}

\address{%
Jesse Wheeler\\
University of Michigan\\%
Department of Statistics\\ Ann Arbor, Michigan\\
\textit{ORCiD: \href{https://orcid.org/0000-0003-3941-3884}{0000-0003-3941-3884}}\\%
\href{mailto:jeswheel@umich.edu}{\nolinkurl{jeswheel@umich.edu}}%
}

\address{%
Aaron A. King\\
University of Michigan and Santa Fe Institute\\%
Department of Ecology and Evolutionary Biology\\ Ann Arbor, Michigan\\
\textit{ORCiD: \href{https://orcid.org/0000-0001-6159-3207}{0000-0001-6159-3207}}\\%
\href{mailto:kingaa@umich.edu}{\nolinkurl{kingaa@umich.edu}}%
}

\address{%
Edward L. Ionides\\
University of Michigan\\%
Department of Statistics\\ Ann Arbor, Michigan\\
\textit{ORCiD: \href{https://orcid.org/0000-0002-4190-0174}{0000-0002-4190-0174}}\\%
\href{mailto:ionides@umich.edu}{\nolinkurl{ionides@umich.edu}}%
}

\end{article}

\end{document}